\documentstyle[11pt,newpasp,twoside,epsf]{article}
\def\edcomment#1{\iffalse\marginpar{\raggedright\sl#1\/}\else\relax\fi}
\reversemarginpar

\begin{document}

\title{Decametric AGNs: FIRST and NVSS maps and radio spectra}

\author{O.V. Verkhodanov, N.V.Verkhodanova}
\affil{Special Astrophysical Observatory, Nizhnij Arkhyz, Russia, 369167}
\author{H. Andernach}
\affil{Depto.\,de Astronom\'\i a, Apdo.\ Postal 144, Univ.\ Guanajuato, Mexico}

\begin{abstract}
Radio sources from the decametric UTR--2 catalog were
cross-identified with other radio catalogs at higher
frequencies.
We used the CATS database to extract all sources within the UTR beam size
($\sim$40$'$) to find candidate radio identifications.
Using the least squares method we fitted the spectrum of each source
with one of a set of curves.
We extracted NVSS and FIRST radio images for the
radio-identified sources, and looked for a possible relation between
size and spectral index.
\end{abstract}

%
A radio survey obtained with the UTR telescope (Kharkov, Ukraine)
at frequencies 10--25\,MHz has resulted in a catalog of 1822 sources
(Braude et al.\ 1978--1994; {\tt www.ira.kharkov.ua/UTR2}).
Covering about 30\% of the sky north of $-$13\deg\ declination,
this survey is presently the lowest-frequency source catalog
of its size, and thus provides an ideal basis to study the little known
optical identification content of sources selected at decametric
frequencies.
The rather large uncertainties of UTR positions ($\sim$0.7\deg) require
an iterative process for finding radio counterparts at successively
higher frequencies (and thus higher positional accuracy). In this we aided
ourselves by selecting previously cataloged sources from the
CATS database (Verkhodanov et al.\ 1997) in a box of
RA$\times$DEC~=~40$'\times$40$'$ centred on the nominal UTR position.
The ``raw'' spectra given by these fluxes were refined using
computer charts of source locations around UTR positions.
All counterparts from TXS, GB6 and PMN within circles of 1$'$ radius were
considered one source.  Groups of sources lying further apart
were assigned separate spectra, each with the UTR flux as upper limit.

We were able to fit spectra for all but 7 of the 2314 radio counterparts to
UTR sources. Fits were either straight (S), convex (C$^{-}$),
or concave (C$^{+}$) curves in the lg\,$\nu$--lg\,S plot.
The distribution of radio source spectra among the various spectral types
is given in Table~1.
The resulting catalog (Verkhodanov et al.\ 2000) includes information
from a large number of electronically available catalogs of radio,
infrared, optical and X--ray sources.

\begin{table}[h!]
\caption{Distribution of radio continuum spectral types of
2307 radio counterparts to UTR sources, where X=log$_{10}$(frequency/MHz),
and Y=log$_{10}$(flux density in Jy)}
\begin{center}
\begin{tabular}{llrr}
\tableline
Spectral class & Fitting function & N & \% \\
\tableline
Straight (S)        & $Y = +A+B*X                   $ &  894 &  39 \\
Convex   ($C^{+}$)  & $Y = +A{\pm}B*X-C*X^2         $ &  184 &   8 \\
Concave  ($C^{-}$)  & $Y = +A-B*X+C*X^2             $ & 1150 &  50 \\
~~~~~~~~~~~~~~~~or  & $Y = {\pm}A{\pm}B*X+C*EXP(-X) $ &   79 &   3 \\
\tableline
\tableline
\end{tabular}
\end{center}
\end{table}

\begin{table}
\caption{Flux and size distribution of UTR counterparts in FIRST and
NVSS: median value for various ranges of
spectral index. $N$ is the number of objects.}
\begin{center}
\begin{tabular}{rrccrcc}
\tableline
\multicolumn{1}{c}{Sp.\,index} & \multicolumn{3}{c}{~~~~~~median FIRST} &
		  \multicolumn{3}{c}{~~~~~~~median NVSS} \\
 range~~~~~  & $N$ & flux (mJy) & size ($''$) & $N$ & flux (mJy) & size ($''$)  \\
\tableline
$-0.9\div-1.0$& 88  &   514  &  26   & 183 &  574 &  24  \\
$-1.0\div-1.1$& 93  &   363  &  19   & 168 &  365 &  22  \\
$<-$1.1       & 75  &   193  &  12   & 185 &  212 &  17  \\
\tableline
\tableline
\end{tabular}
\end{center}
\end{table}

The majority of UTR sources (97\%) have an identification
with NVSS objects (Condon et al., 1998)
(a total of 2253 IDs),
and all UTR objects with
$\delta>30\deg$ (1143 objects) have IDs in FIRST (White et al., 1997).
552 sources were resolved into components either in
the FIRST ($5\arcsec$ beam) or NVSS ($15\arcsec$ beam) catalogs
within a circle of 60$\arcsec$.

We extracted NVSS and VLA maps of sources with steep and straight power 
law spectra, and grouped them according to their spectral slope:
(1) $-0.9>\alpha>-1.0$,
(2) $-1.0>\alpha>-1.1$ and (3) $\alpha<-1.1$ ($S\propto\nu^{\alpha}$).
NVSS objects were taken outside of the
Galactic plane ($|b| > 15\deg$), while all the FIRST sources
have $|b| > 20\deg$. 536 NVSS objects and
256 FIRST objects were selected.
Parameters of source samples are provided in Table~2.
%
Surprisingly, the median source size decreases
with steepening radio spectrum.  However, the dispersion in size is so large 
that there is no significant correlation between spectral index and source size.



\begin{references}
\reference
Braude S.Ya. et al. 1978,1979,1981,1985,1994, Ap\&SS, {\bf 54},37; {\bf 64},73;
{\bf 74},409; {\bf 111},1; {\bf 213},1
\reference
Condon J.J., Cotton W.D., Greisen E.W., Yin Q.F.,
   et al.\ 1998, AJ 115, 1693
\reference
 Verkhodanov O.V., Trushkin S.A., Andernach H., \& Chernenkov V.N. 1997,
     in "Astronomical Data Analysis Software and Systems VI",
     eds.\ G.Hunt \& H.E.Payne, ASP Conf.\ Ser.\ {\bf 125}, 322
     (astro-ph/9610262)
\reference
 Verkhodanov O.V., Andernach H., Verkhodanova N.V. 2000,
     Bull.\,Spec. Astroph. Obs. {\bf 49}, 53 (astro-ph/0008431)
\reference
White R.L., Becker R.H., Helfand D.J., \& Gregg M.D.\ 1997, ApJ 475, 479 
\end{references}
\end{document}